\documentclass[preprint2]{aastex}

\shorttitle{}
\shortauthors{Richer et al.}

\begin{document}


\title{The Acceleration of the Nebular Shells in Planetary Nebulae \\ in the Milky Way Bulge\footnote{The observations reported herein were obtained at the Observatorio Astron\'omico Nacional in the Sierra San Pedro M\'artir (OAN-SPM), B. C., Mexico.}}


\author{Michael G. Richer, Jos\'e Alberto L\'opez, Margarita Pereyra, Hortensia Riesgo,\\ Mar\'\i a Teresa Garc\'\i a-D\'\i az}
\affil{OAN, Instituto de Astronom\'\i a, Universidad Nacional Aut\'onoma de M\'exico, \\ P.O. Box 439027, San Diego, CA 92143}
\email{\{richer, jal, mally, hriesgo, tere\}@astrosen.unam.mx}

\and

\author{Sol-Haret B\'aez}
\affil{Facultad de F\'\i sica e Inteligencia Artificial, Universidad Veracruzana, \\ Circuito G. Aguirre Beltr\'an s/n, Zona Universitaria, Xalapa, Veracruz, M\'exico}
\email{solharet@hotmail.com}



\begin{abstract}

We present a systematic study of line widths in the [\ion{O}{3}]$\lambda$5007 and H$\alpha$ lines for a sample of 86 planetary nebulae in the Milky Way bulge based upon spectroscopy obtained at the \facility{Observatorio Astron\'omico Nacional in the Sierra San Pedro M\'artir (OAN-SPM)} using the Manchester Echelle Spectrograph.  The planetary nebulae were selected with the intention of simulating samples of bright extragalactic planetary nebulae.  We separate the planetary nebulae into two samples containing cooler and hotter central stars, defined by the absence or presence, respectively, of the \ion{He}{2}\,$\lambda$6560 line in the H$\alpha$ spectra.  This division separates samples of younger and more evolved planetary nebulae.  The sample of planetary nebulae with hotter central stars has systematically larger line widths, larger radii, lower electron densities, and lower H$\beta$ luminosities.  The distributions of these parameters in the two samples all differ at significance levels exceeding 99\%.  These differences are all in agreement with the expectations from hydrodynamical models, but for the first time confirmed for a homogeneous and statistically significant sample of galactic planetary nebulae.  We interpret these differences as evidence for the acceleration of the nebular shells during the early evolution of these intrinsically bright planetary nebulae.  As is the case for planetary nebulae in the Magellanic Clouds, the acceleration of the nebular shells appears to be the direct result of the evolution of the central stars.  

\end{abstract}





\section{Introduction}

The interacting stellar winds model of \citet{kwoketal1978} provides a general framework for understanding planetary nebulae.  Within this framework, a variety of theoretical and numerical studies of the hydrodynamics have been undertaken to understand and predict the kinematic properties of planetary nebulae.  Our modern view of the kinematics of planetary nebulae was established in the early 1990's \citep[e.g.,][]{kahnwest1985, breitschwerdtkahn1990, kahnbreitschwerdt1990, mellema1994}.  These studies showed that the AGB envelope is accelerated in two phases.  The first phase is a result of the shock wave initiated by the ionization front that sweeps through the AGB envelope as the central star's temperature increases.  Then, as the central star's wind energy increases, it drives a pressure-driven central bubble that accelerates the AGB  envelope through ram pressure.  More recent numerical work that attempts to include more realistic AGB evolution confirms these basic results \citep[e.g.,][]{villaveretal2002, perinottoetal2004}.  Eventually, the central star's wind ceases and the inner part of the AGB envelope backfills towards the central star while the outer part maintains its momentum-driven evolution \citep[e.g.,][]{garciaseguraetal2006}.  

These models have been used extensively to interpret observational studies of the kinematics of many individual planetary nebulae.  Unfortunately, to date,  systematic, homogeneous studies of planetary nebula populations to compare with these theoretical efforts are scarce.  \citet{heap1993} found a correlation between nebular expansion velocity and the terminal velocity of the central star wind.  More recently, \citet{medinaetal2006} claimed a correlation between expansion velocity and age indicators such as density and central star temperature.  However, both of these studies suffer from heterogeneity, at least regarding object selection.  Even regarding planetary nebulae with Wolf-Rayet (WR) central stars, the precise role that the winds from the central stars play in the kinematics of the nebular shells is not entirely clear \citep{gesickietal2006, medinaetal2006}.

The studies of the kinematics of planetary nebulae in the Magellanic Clouds are perhaps the most systematic \citep{dopitaetal1985, dopitaetal1988}.  These studies find correlations between expansion velocity, excitation class, and nebular density, suggesting that the properties of the central star dominate the nebular evolution.  As they point out, however, their results are not general since they have been found only for a particular population of (bright) planetary nebulae.  

Therefore, there remains a need for a coherent population study in a different galactic environment.  Here, we present a study of the kinematics for a large sample of planetary nebulae in the bulge of the Milky Way (Bulge).  These objects were selected in a homogeneous way, with the hope of simulating populations of bright extragalactic planetary nebulae in bulge-like systems (bulges of spiral galaxies, dwarf spheroidals, and elliptical galaxies).  The observations were all made with the same instrument and methodology.  Likewise, the data reduction and analysis is homogeneous and entirely independent of model parameters.  Details are given in Section 2.  We find that the expansion properties do indeed depend upon the evolutionary stage of the central star, with planetary nebulae hosting hotter central stars having larger expansion velocities (Section 3).  
We also find that other parameters that should depend upon age, the nebular size, density, and H$\beta$ luminosity, also differ significantly between the two groups (Section 4).  We compare these properties with theoretical models and find generally good agreement (Section 5).  We present our conclusions in Section 6.

\section{Observations and Analysis}

\subsection{The Planetary Nebula Sample}

We present the sample of Bulge planetary nebulae that we observed in Table \ref{table_objects}.  There are 86 objects in total, drawn from existing spectroscopic surveys \citep{allerkeyes1987, webster1988, cuisinieretal1996, ratagetal1997, cuisinieretal2000, escuderocosta2001, escuderoetal2004, exteretal2004, gornyetal2004}.

We selected this sample hoping to simulate a sample of bright extragalactic planetary nebulae in bulge-like systems.  Our selection criteria were refined somewhat over the duration of the observations, but eventually converged to (i) a position within $10^{\circ}$ of the galactic centre, (ii) a large observed, reddening-corrected H$\beta$ flux, nominally $\log I(\mathrm H\beta) > -12.0$\,dex, (iii) a large [\ion{O}{3}]$\lambda 5007/\mathrm H\beta$ ratio, usually exceeding a value of 6, and (iv) the existence of low resolution spectroscopy in which the electron temperature may be determined from the [\ion{O}{3}]$\lambda 4363/5007$ ratio.  Usually, the systemic radial velocity exceeds 30 km/s in an attempt to exclude disk objects towards the Bulge.  We imposed no explicit limit on the size of the objects.  Nor did we impose an upper limit upon the flux (H$\beta$ or 6\,cm).  In practice, however, a flux limit is usually imposed by the surveys from which we selected our objects.

\begin{figure*}
\includegraphics[scale=0.6,angle=-90]{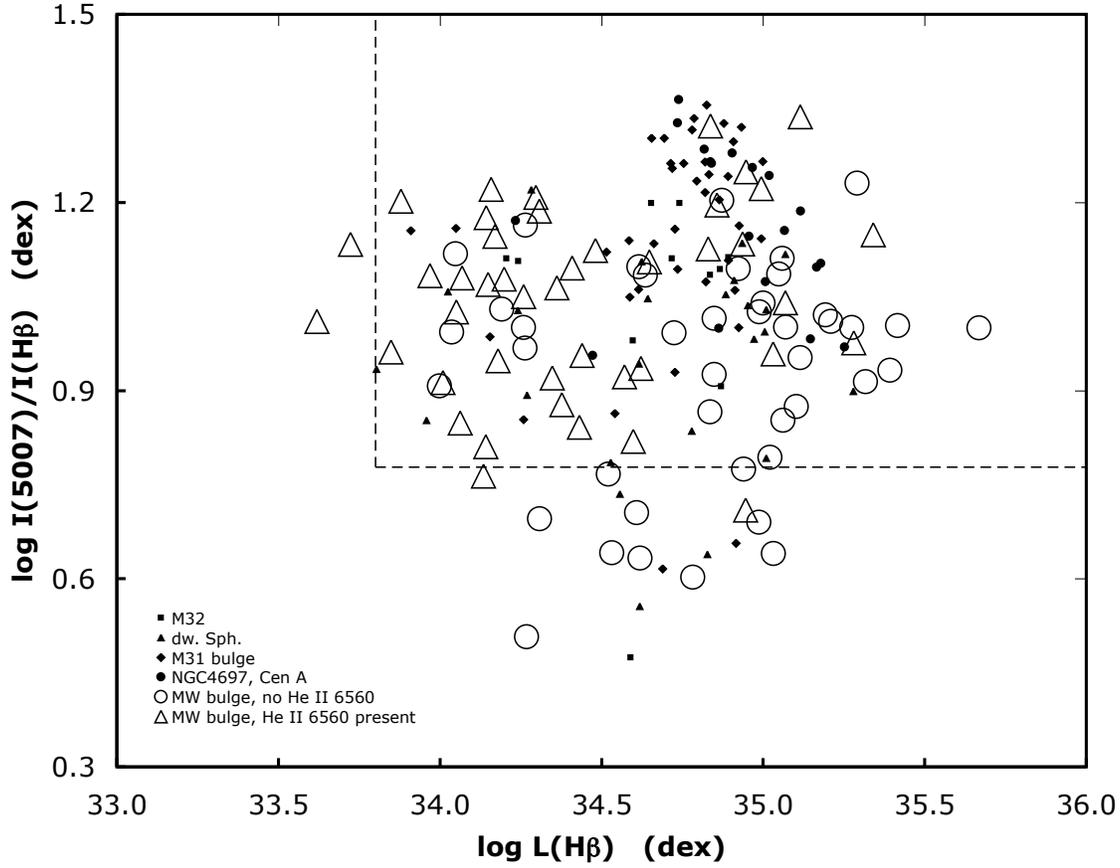}
\caption{The selection criteria for our sample of Bulge planetary nebulae were designed to yield a sample that mimics the populations of bright extragalactic planetary nebulae in bulge-like systems.  The extragalactic planetary nebulae are shown in solid symbols while the Bulge planetary nebulae, divided according to the presence or absence of \ion{He}{2}$\lambda$6560 are shown with open symbols.  The dotted lines indicate the limits we chose for the H$\beta$ luminosity and [\ion{O}{3}]$\lambda 5007/\mathrm H\beta$ ratio were motivated by the range of these parameters among extragalactic planetary nebulae with spectroscopic observations \citep{jacobyciardullo1999, richeretal1999, walshetal1999, dudziaketal2000, rothetal2004, mendezetal2005, zijlstraaetal2006, goncalvesetal2007, richermccall2008}.  \label{fig_crit_xgal}}
\end{figure*}

Figure \ref{fig_crit_xgal} explains the logic of our lower limits on the H$\beta$ flux and the [\ion{O}{3}]$\lambda 5007/\mathrm H\beta$ ratio.  Basically, bright extragalactic planetary nebulae in ellipticals, dwarf spheroidals, and the bulge of M31 fall within these limits (see Figure caption).  Restricting the sample to objects with large [\ion{O}{3}]$\lambda\lambda 5007/\mathrm H\beta$ ratios will exclude planetary nebulae very early in their evolution when their central stars are coolest.  Our limit on the H$\beta$ luminosity does not restrict the planetary nebulae to contain central stars that are on the horizontal portion of their evolutionary track, but also includes the initial fading following the extinction of nuclear reactions.  

\subsection{Observations and Data Reductions}

We acquired our observations during eight observing runs spanning the period from 2003 June to 2007 August at the Observatorio Astron\'omico Nacional in the Sierra San Pedro M\'artir, Baja California, Mexico (OAN-SPM).  More details of the observations will be provided in a forthcoming paper.  

High resolution spectra were obtained with the Manchester echelle spectrometer \citep[MES-SPM; ][]{meaburnetal1984, meaburnetal2003}.  The MES-SPM is a long slit, echelle spectrometer that has no cross-dispersion. Instead, narrow-band filters were used to isolate orders 87 and 114 containing the H$\alpha$ and [\ion{O}{3}]$\lambda 5007$ emission lines, respectively.  A 150\,$\mu$m wide slit was used for the observations, resulting in a slit $1\farcs9$ wide on the sky.  Coupled with a SITe $1024\times 1024$ CCD with 24\,$\mu$m pixels binned $2\times 2$, the resulting spectral resolutions were approximately 0.077\,\AA/pix and 0.100\,\AA/pix at [\ion{O}{3}]$\lambda 5007$ and H$\alpha$, respectively (equivalent to 11\,km/s for 2.6\,pix FWHM). The spectra were calibrated using exposures of a ThAr lamp taken immediately before or after every object exposure.  The internal precision of the arc lamp calibrations is better than $\pm 1.0$\,km/s. 

Usually, only one deep spectrum was obtained in each of the [\ion{O}{3}]$\lambda 5007$ and H$\alpha$ filters.  The exposure times varied depending upon the brightness of the object.  At most, the exposure times were of 30 minutes duration.  If this was expected to produce saturation, shorter exposure times were used.  The exposure time for the H$\alpha$ spectrum was chosen to achieve a signal level similar to that in the deep [\ion{O}{3}]$\lambda 5007$ spectrum, based upon the intensities of these lines observed in the low resolution spectra, but was also limited to a maximum of 30 minutes duration.

All of the planetary nebulae observed are resolved (see Table \ref{table_objects}).  In all cases, the slit was centered on the object.  Usually, the slit was oriented in the north-south direction.  

The spectra were reduced using the twodspec and specred packages of the Image Reduction and Analysis Facility\footnote{IRAF is distributed by the National Optical Astronomical Observatories, which is operated by the Associated Universities for Research in Astronomy, Inc., under contract to the National Science Foundation.} (IRAF).  The procedure for data reduction followed that outlined in \citet[][Appendix B]{masseyetal1992} for long slit spectroscopy.  The object spectra were edited of cosmic rays.  Then, a nightly mean bias image was subtracted from each object spectrum.  Next, the arc lamp spectra were used to map positions of constant wavelength.  These maps were then used to rectify the object spectra so that lines of constant wavelength fell exactly along the columns, a process that simultaneously applies a wavelength calibration.  Finally, wavelength-calibrated, one-dimensional spectra were extracted for each object.  No flux calibration was performed.  

\subsection{Kinematic parameters}

We shall present all of the one-dimensional line profiles in a subsequent publication.  For that reason, we do not present the line profiles here.  Instead, we turn to an explanation of the analysis of the kinematics.

The one-dimensional spectra were analyzed using a locally-implemented software package \citep[INTENS;][]{mccalletal1985} to determine the radial velocity, flux, and profile width (FWHM; full width at half maximum intensity) as well as the uncertainties in these parameters.  This software fits the emission line profile with a sampled gaussian function and models the continuum as a straight line.  Thus, this analysis assumes that the lines have a gaussian shape and that they are superposed on a flat continuum.  In the case of the H$\alpha$ line, the \ion{He}{2}\,$\lambda$6560 line may also be present.  In this case, a fit is made simultaneously to both lines and the continuum.    

\begin{figure}
\begin{center}
\includegraphics[scale=0.31,angle=-90]{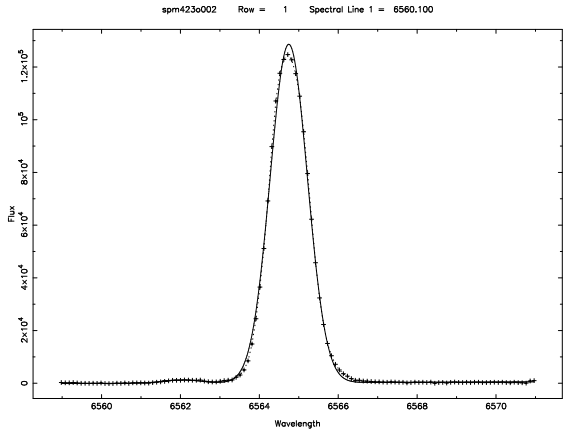}
\end{center}
\caption{The H$\alpha$ line profile for M 2-16 is an example of our spectra.  The crosses are the data and the solid line is the fit.  The faint line on the blue wing of H$\alpha$ is the \ion{He}{2}\,$\lambda$6560 line and has an intensity of 0.9\% that of H$\alpha$.  For the H$\alpha$ line, the S/N in the flux is 200, which is approximately the median value for our sample.  \label{fig_m216ha}}
\end{figure}

While the assumption of a gaussian line shape is reasonable for bright extragalactic planetary nebulae, because they are spatially unresolved, it might seem odd for spatially-resolved objects.  Nonetheless, for the large majority of the objects, the one-dimensional line profiles are usually not double-peaked and the deviations from a gaussian shape usually represent less than 10\% of the total flux \citep[see Fig. \ref{fig_m216ha} for an example;][]{richeretal2009}.  

Table \ref{table_objects} presents the observed line widths (FWHM) and their uncertainties for each object in both H$\alpha$ and [\ion{O}{3}]$\lambda 5007$.  The uncertainties in the line widths are the formal uncertainties (one sigma) from fitting a sampled gaussian function to the line profile.  In order to derive the intrinsic line widths, the observed line widths must be corrected for several effects, all of which are assumed to contribute to the observed line width in quadrature.  The effects that broaden the true, intrinsic profile are instrumental ($\sigma_{inst}$), thermal ($\sigma_{th}$), and fine structure ($\sigma_{fs}$) broadening,

\begin{equation}
\sigma^2_{obs} = \sigma^2_{true} + \sigma^2_{inst} + \sigma^2_{th} + \sigma^2_{fs}\ .
\end{equation}

\noindent  The first term, $\sigma^2_{true}$, is the true, intrinsic line width resulting from the kinematics of the planetary nebula.  The instrumental profile has a measured FWHM of 2.5-2.7 pixels.  We adopt a FWHM of 2.6 pixels for all objects, which amounts to about 11\,km/s (FWHM).  We compute the thermal broadening from the usual formula \citep[][eq. 2-243]{lang1980}, adopting rest wavelengths of 6562.83\AA\ and 5006.85\AA\ for H$\alpha$ and [\ion{O}{3}]$\lambda 5007$, respectively, and assuming no turbulent velocity.  The thermal broadening (FWHM) amounts to 0.47\AA\ (21.4\,km/s) and 0.089\AA\ (5.3\,km/s) for H$\alpha$ and [\ion{O}{3}]$\lambda 5007$, respectively.  The fine structure broadening was taken to be $\sigma_{fs} = 3.199$\,km/s for H$\alpha$ and zero for [\ion{O}{3}]$\lambda 5007$ \citep{garciadiazetal2008}.  

The interpretation of the resulting line width, $\Delta V$, 

\begin{equation}
\Delta V = 2.3556\sigma_{true}, 
\end{equation}

\noindent is not necessarily simple.  In the case of an expanding, spherical shell that is spatially-unresolved, the line width corresponds to the expansion velocity.  If the velocity field or matter distribution is more complicated or if the object is resolved spatially, then the line width is a luminosity-weighted velocity width for the mass projected within the spectrograph slit.  

The velocity width that we measure should typically exceed the luminosity-weighted line width for the entire object.  The spectrograph slit was centered on each object and all of the objects are wider than the slit (Table \ref{table_objects}).  Therefore, matter near the edges of the objects is excluded and this matter is likely to have the lowest projected velocity along the line of sight.  Since the matter at the edges of the objects would contribute  low systematic velocities for each object, the line profile we measure for the matter included within the slit will be slightly larger than the true mass-weighted line width.  These arguments are supported by the results presented by \citet{gesickizijlstra2000} and \citet{rozasetal2007}.  Their simulations of thin, expanding, spherical shells indicate that the line widths we measure may over-estimate the integrated line widths for the entire objects by up to approximately 15\%, but that the exact amount will depend upon the fraction of the object covered by the slit.  

Fortunately, the discussion that follows does not depend upon any interpretation of the line width.  Since it is clear, however, that this line width will be similar to twice the expansion velocity, Table \ref{table_objects} presents half of the line width in velocity units for each object, i.e.,   

\begin{equation}
\Delta V_{0.5} = 0.5\Delta V = 1.1778\sigma_{true}\ .
\end{equation}  

\section{Nebular Kinematics versus the Evolutionary State of the Central Star}

For about half of the objects in our sample, the \ion{He}{2}\,$\lambda$6560 line is present in the H$\alpha$ spectra.  Fig. \ref{fig_m216ha} presents M 2-16 as an example.  In practice, we can detect \ion{He}{2}\,$\lambda$6560 provided its intensity is at least 0.4\% that of H$\alpha$ and a signal-to-noise of at least 50 is achieved for the H$\alpha$ line.  For comparison with theoretical models, this limit is predicted to occur for an effective temperature of about $9\times 10^4$\,K according to the hydrodynamical models presented in \citet{schonberneretal2007}, if we use the the atomic data for hydrogen and helium in \citet{osterbrock1989}.  Table \ref{table_objects} indicates the objects for which the \ion{He}{2}\,$\lambda$6560 line was observed in our H$\alpha$ spectra.  

Obviously, the objects in which the \ion{He}{2}\,$\lambda$6560 line appears are those with the hotter central stars.  Therefore, we may separate our sample into two samples with cooler and hotter central stars according to whether the \ion{He}{2}\,$\lambda$6560 line is observed.  Given our selection criteria, this separation into samples with cooler and hotter central stars is equivalent to a separation into samples in different stages of evolution.  The planetary nebulae with the hotter central stars are in a more advanced stage of evolution than those with the cooler central stars.  

\begin{figure*}
\begin{center}
\includegraphics[scale=0.32,angle=-90]{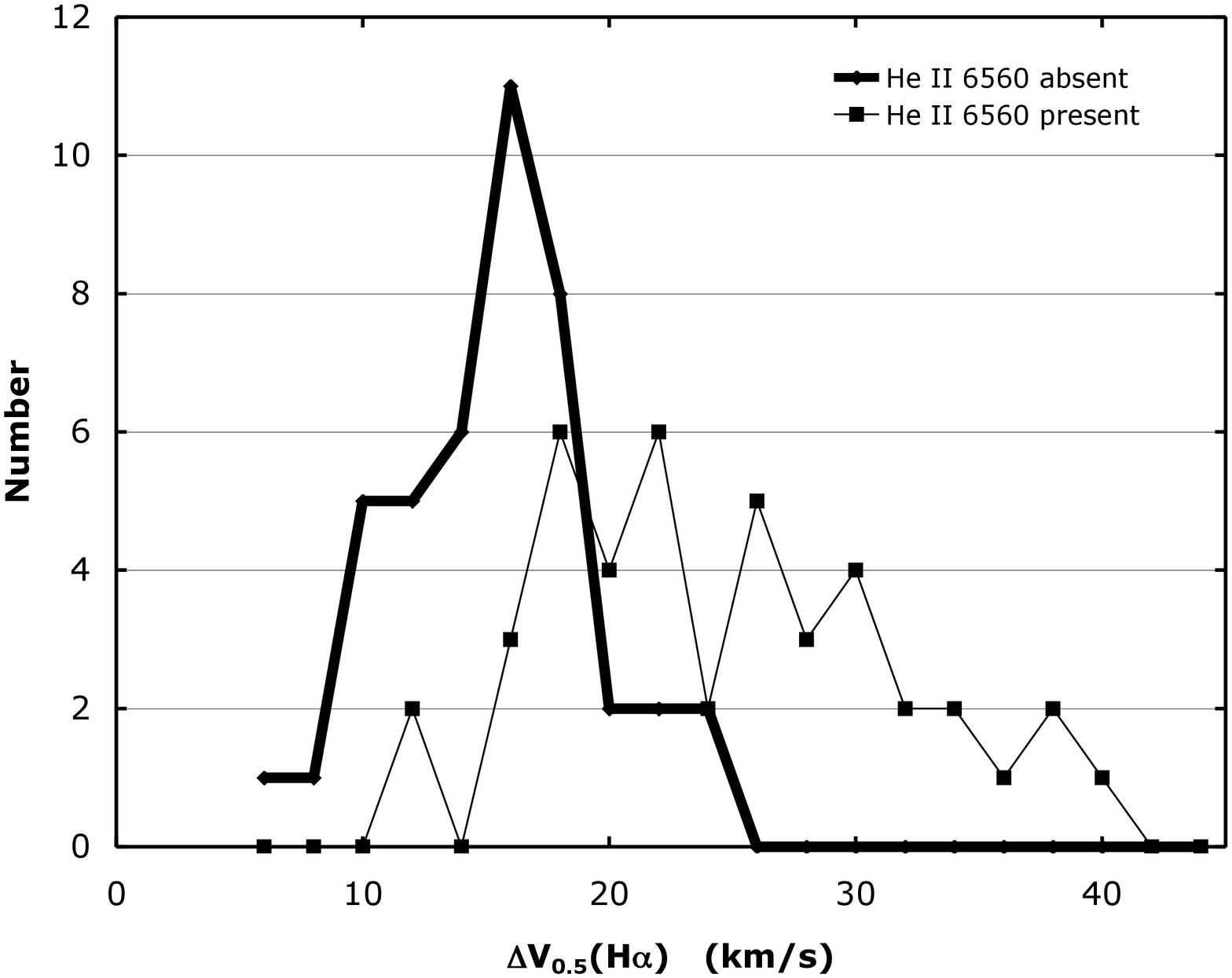}\,\quad\,
\includegraphics[scale=0.32,angle=-90]{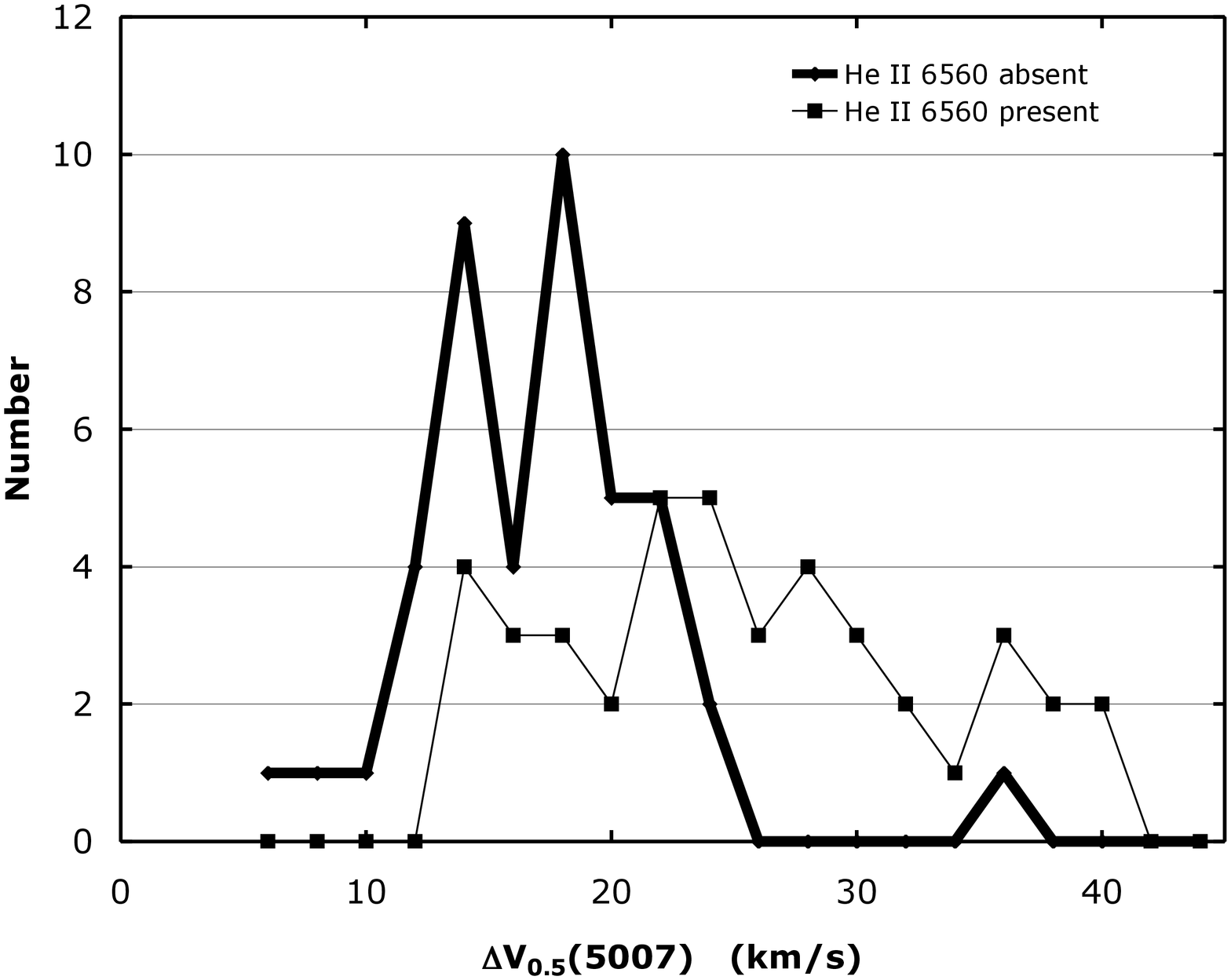}
\end{center}
\caption{We find distributions of line widths that are very different if we divide our objects according to whether the \ion{He}{2}$\lambda$6560 line is absent or present.  The distribution of H$\alpha$ line widths is on the left while the distribution of [\ion{O}{3}]$\lambda$5007 line widths is on the right.    The probability of drawing such different distributions (\ion{He}{2}$\lambda$6560 absent or present) at random from the same parent population is less than $1.3\times 10^{-7}$ (Table \ref{table_statistics}).  \label{fig_dist_lw}}
\end{figure*}

Fig. \ref{fig_dist_lw} presents histograms of the line width ($\Delta V_{0.5}$) distributions for the two samples of planetary nebulae.  Clearly, the distributions appear different.  The sample of planetary nebulae with the cooler central stars have predominantly low line widths, while the sample with the hotter central stars have a much broader distribution of line widths that extends to substantially higher values.  

A variety of statistical tests may be applied to these distributions to determine whether it is probable that they may result from a single parent distribution.  Since we do not know a priori the form of the distribution that the line widths should have, non-parametric statistical tests are most appropriate.  Perhaps, the best-known of these tests is the Kolmogorov-Smirnov two-sample test to the cumulative distribution functions of the line widths.  However, a more powerful non-parametric test is the Wilcoxon-Mann-Whitney U-test \citep[e.g.,][\S 5.4.3]{walljenkins2003}.  This is a rank test in which the line widths from the two samples are combined into a single sample and ranked.  Then, the rankings of the line widths for the two samples are compared with a uniform distribution of rankings.  

The U-test indicates that the probability of drawing the two samples of line widths from the same parent distribution is only $1.3\times 10^{-7}$.  The  nebular shells surrounding hotter central stars are expanding systematically more rapidly than their counterparts surrounding cooler central stars.  Since all of these hotter central stars were necessarily cooler in the past, the simplest conclusion is that the nebular shells in bright planetary nebulae are accelerated as the star gets hotter.  

\section{Other Properties versus the Evolutionary State of the Central Star}

Considering that the nebular shells of bright planetary nebulae are accelerated as the central star gets hotter, one might expect other properties of the two samples of planetary nebulae to differ.  Essentially, characteristics that vary systematically with the age or evolutionary stage of the planetary nebula should differ between the two samples.

The most obvious characteristic to consider is the nebular size.  The cooler central stars should be found in younger planetary nebulae that have had less time to expand, and should be systematically smaller than those surrounding the hotter central stars.  There are no systematic measurements of the diameters of the planetary nebulae in our sample, so we measured nebular diameters from our deep H$\alpha$ spectra.  To determine the diameters, we collapsed the spectra along the wavelength axis to produce one-dimensional spatial profiles.  We then measured the nebular sizes at 50\% and 10\% of the peak intensity.  These diameters are found in Table \ref{table_objects}.

\begin{figure}
\begin{center}
\includegraphics[scale=0.31,angle=-90]{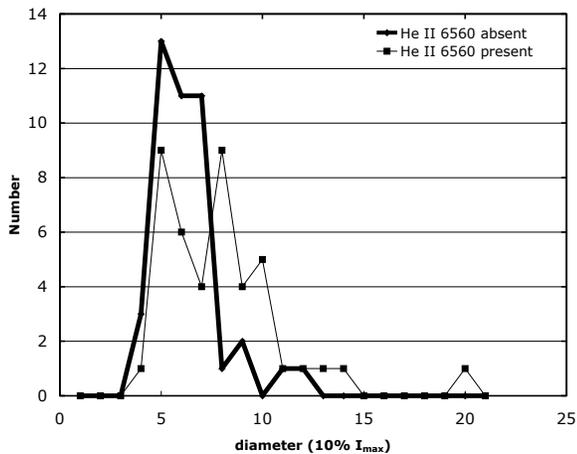}
\end{center}
\caption{The distributions of diameters at 10\% of peak intensity (or 50\% of peak) are very different for the Bulge planetary nebulae divided according to whether the \ion{He}{2}$\lambda$6560 line is absent or present.  The probability of drawing such different distributions of diameters at random from the same parent population is $7.0\times 10^{-4}$ (Table \ref{table_statistics}).  \label{fig_dist_diam10}}
\end{figure}

The distributions of diameters at 10\% of the peak flux are shown in Fig. \ref{fig_dist_diam10}.  Again, it is clear that the sample with the hotter central stars has a distribution of diameters that tends to larger sizes than the sample of planetary nebulae with the cooler central stars.  Applying the U-test, we obtain a probability of only $7\times 10^{-4}$ that the two samples arise from a single parent population.  Hence, the nebulae surrounding the hotter central stars are larger than those surrounding the cooler, less evolved, central stars.  

Conceivably, the larger sizes of the nebulae around hotter central stars might artificially bias their line widths to larger values because the fixed spectrograph slit covers a smaller fraction of these larger nebulae.  As shown by \citet{gesickizijlstra2000} and \citet{rozasetal2007}, this might cause the line widths for larger nebulae to be over-estimated.  To test whether this is the case, we estimated how much the line widths could be over-estimated based upon the results presented in Table 1 of \citet{gesickizijlstra2000}.  We then divided the line widths of all objects in Table \ref{table_objects} ($\Delta V_{0.5}$) by this correction and re-calculated the probability that the two samples arise from a single parent population.  The probability increases, but to only $2.5\times 10^{-7}$ (considering the [\ion{O}{3}]$\lambda$5007 line widths), so the conclusion that the two distributions are statistically very different is not affected.  

\begin{figure}
\begin{center}
\includegraphics[scale=0.3,angle=-90]{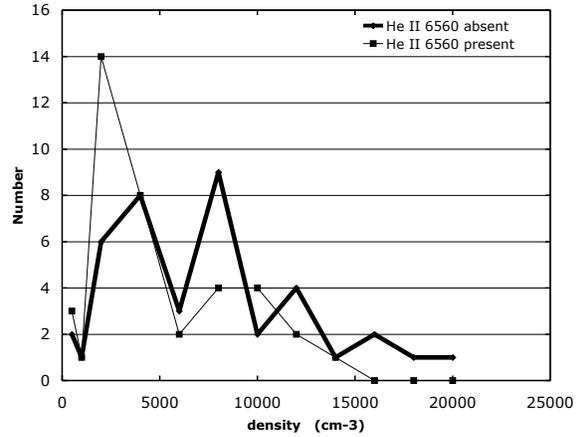}
\end{center}
\caption{The distributions of densities are very different for the Bulge planetary nebulae hosting cooler and hotter central stars (\ion{He}{2}$\lambda$6560 absent and present, respectively).  The probability of drawing these distributions of densities at random from the same parent population is $8.0\times 10^{-3}$ (Table \ref{table_statistics}).  \label{fig_dist_density}}
\end{figure}

The nebular density is also expected to evolve, decreasing as the nebular volume increases.  Table \ref{table_objects} lists the electron densities determined from the [\ion{S}{2}]$\lambda\lambda$6716,6731 ratio as measured in low resolution spectroscopy from the literature.  When there was more than one measurement, the measurements were averaged.  The distributions of nebular density in the two samples are shown in Fig. \ref{fig_dist_density}.  In this case, the difference is more subtle, but the planetary nebulae with the hotter central stars tend to have lower densities.  Applying the U-test, we find a probability of $8\times 10^{-3}$ that the two samples arise from the same parent population.  Therefore, the planetary nebulae with hotter central stars have systematically lower electron densities.  

Finally, we compare the H$\beta$ luminosities of the two samples.  We expect all of the planetary nebulae without \ion{He}{2}$\lambda$6560 emission to contain central stars on the horizontal portion of their evolutionary tracks.  It is unlikely that any of these objects contain central stars that are fading towards the white dwarf regime since virtually all of the central stars should be sufficiently massive to achieve temperatures in excess of $10^5$\,K before fading, by which time emission from He$^{2+}$ ions should be evident.  On the other hand, the sample of planetary nebulae with hotter central stars should contain objects whose central stars are fading towards the white dwarf regime.  Therefore, we would expect that the H$\beta$ luminosities of the two samples could differ, with the sample with the cooler central stars having higher luminosities.  

The H$\beta$ luminosities for the Bulge planetary nebulae in Table \ref{table_objects} and Fig. \ref{fig_crit_xgal} are not very accurate for a variety of reasons.  First, the observed H$\beta$ fluxes are usually those measured through a spectrograph slit and not from photometry.  Likely, the H$\beta$ fluxes are underestimated for many Bulge planetary nebulae.  Second, these fluxes must be corrected for reddening.  The reddening may include instrumental or observational effects, such as differential atmospheric refraction, and so may over-estimate the true reddening.  These reddenings have been used to correct the H$\beta$ fluxes for extinction.  Third, the individual distances for the planetary nebulae are unknown.  To compute the H$\beta$ luminosities in Table \ref{table_objects}, we adopted a common distance of 7.5\,kpc.  In all likelihood, the mean distance for the sample will be less than this and it is probably this assumption that leads to larger H$\beta$ luminosities than are observed in extragalactic planetary nebulae.  

\begin{figure}
\begin{center}
\includegraphics[scale=0.31,angle=-90]{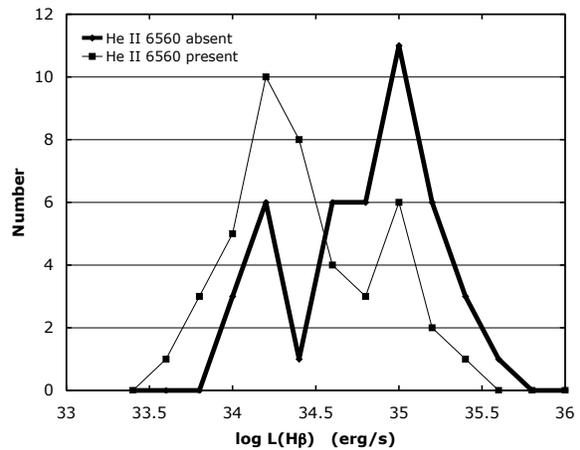}
\end{center}
\caption{As was found for the line widths, diameters, and densities, the distributions of H$\beta$ luminosities are very different for the Bulge planetary nebulae hosting cooler and hotter central stars (\ion{He}{2}$\lambda$6560 absent and present, respectively).  The probability of drawing these distributions of luminosities at random from the same parent population is $2.2\times 10^{-4}$ (Table \ref{table_statistics}).  \label{fig_dist_hb}}
\end{figure}

Fig. \ref{fig_dist_hb} presents histograms of the H$\beta$ luminosity distributions for the two samples, divided according to the presence or absence of \ion{He}{2}$\lambda$6560 emission.  The H$\beta$ luminosities are taken from Table \ref{table_objects}.  Clearly, the distribution for the sample with the cooler central stars extends to higher H$\beta$ luminosities.  The mean luminosities for the two samples differ by 0.36\,dex, or 0.90\,mag.  The U-test indicates that the probability that two distributions arise from the same parent population is only $2.2\times 10^{-4}$.  Therefore, the planetary nebulae with the cooler central stars are brighter, on average, than those with the hotter central stars.  

Summarizing our results, we find significant differences between our samples of planetary nebulae with cooler and hotter central stars.  The planetary nebulae with cooler central stars have smaller line widths, smaller sizes, higher densities, and higher H$\beta$ luminosities.  The differences are all significant at confidence levels exceeding 99\% (Table \ref{table_statistics}).  

\section{Discussion}

At present, the theory of the evolution of planetary nebulae outlined earlier is unable to predict definitive expansion velocities for planetary nebulae or the AGB stars from which they arise.  Observations of AGB stars find that their winds usually have an expansion velocity below 15\,km/s and often below 10\,km/s \citep[e.g.,][]{ramstedtetal2006, lewis1991}.  Observations of planetary nebulae typically find expansion velocities of 15-30\,km/s \citep[e.g.,][]{gesickizijlstra2000, medinaetal2006}.  At least qualitatively, theory may explain the difference:  The ionization front drives a shock wave through the AGB envelope, raising the expansion velocity by $\sim 5-6$\,km/s over that of the AGB envelope \citep{chevalier1997, perinottoetal2004}.  Then, the pressure-driven bubble  expands into the the AGB envelope, sweeping up a dense, inner rim of matter that may increase the expansion velocity significantly, particularly if this rim is able to overtake the density enhancement produced by the shock wave.

The median line width in [\ion{O}{3}]$\lambda$5007 (H$\alpha$) for our sample of Bulge planetary nebulae with cooler central stars is 17.2\,km/s (15.4\,km/s) and the standard deviation of the distribution is 5.1\,km/s (4.0\,km/s).  Since we expect our line widths to be slightly larger than the expansion velocity, the median line width for the sample with the cooler central stars is approximately that expected if the AGB envelopes had expansion velocities of about 10\,km/s, as observations suggest commonly occurs \citep{ramstedtetal2006}.  It is therefore natural to associate these planetary nebulae with models during the stage in which the ionization front is driving a shock wave through the AGB envelopes.  

The median line width in [\ion{O}{3}]$\lambda$5007 (H$\alpha$) for the sample of Bulge planetary nebulae with the hotter central stars is substantially higher, 23.7\,km/s (24.3\,km/s), as is the dispersion in velocities, 7.6\,km/s (7.1\,km/s).  In the models, by the time that the central star has a temperature of $9\times 10^4$\,K, the stellar wind-powered bubble is expanding into the ionized AGB envelope and raising the observed expansion velocities.  Thus, our sample of planetary nebulae with hotter central stars is naturally associated with this second stage of envelope acceleration \citep{villaveretal2002, perinottoetal2004}.  

It is more difficult to compare the differences in size that we find for our two samples with existing theoretical models.  In part, this has to do with how we have determined our nebular sizes.  The main problem is the mismatch in spatial resolution between our observations and one-dimensional theoretical profiles.  Our observations generally include a substantial fraction of the object, while one-dimensional theoretical profiles have infinite spatial resolution.  In spite of this difficulty, the available theoretical profiles indicate that the outermost dense/bright regions increase in size as time progresses \citep{mellema1994, villaveretal2002, perinottoetal2004}, even though the size of the \emph{brightest} structure may not increase monotonically with time \citep{villaveretal2002}.  

Theoretical models indicate that the entire AGB envelopes should become ionized relatively early for the stellar masses of relevance here \citep[$M_i\lesssim 2M_{\odot}$: e.g.,][]{mellema1994, villaveretal2002, perinottoetal2004}.  They are also expected to remain entirely ionized.  Once the envelope is entirely ionized, its H$\beta$ luminosity will decrease with time as a result of dilution \citep[e.g.,][]{osterbrock1989}.  As a result, that we find lower H$\beta$ luminosities for the sample of objects that is more evolved is entirely compatible with theoretical expectations.  

Likewise, theoretical models are compatible with our findings concerning the electron density.  Theoretical models indicate that the densities are higher while the expansion velocities are dominated by the ionization front than later on when the pressure-driven bubble dominates.  However, this agreement is bitter-sweet.  We use densities derived from [\ion{S}{2}] lines.  In principle, these lines from the recombination zone should not exist in the objects with the hotter central stars, since these objects are predicted to have completely ionized their AGB envelopes.  In practice, this emission no doubt arises in density enhancements that are absent from the models, and so the [\ion{S}{2}] densities are likely an upper limit to the typical densities in these objects.  Nonetheless, the models predict a general dilution of the entire AGB envelope and this dilution is probably what gives rise to the decay in [\ion{S}{2}] densities that we observe.

It is not unusual that the line width distributions overlap for our samples of planetary nebulae with cooler and hotter central stars.  As the theoretical models make very clear, the expansion velocity depends upon the AGB envelope expansion velocity and the mass loss rate at the end of the AGB evolution.  Undoubtedly, these properties varied among the stellar progenitors of the planetary nebulae in both samples.  Furthermore, even if they did not, the planetary nebulae in both samples are found in a range of evolutionary states, particularly for the sample with the hotter central stars.  Consequently, the line widths we observe will have varying contributions from the accelerations due to the shock front and the pressure-driven bubble.  Unless the objects are all spherical (unlikely), the different orientations will also broaden the line width distributions.

It is also not unusual that the distributions of line widths differ more than do those for the nebular diameters, densities, and luminosities.  The structure that appears brightest varies with time \citep{villaveretal2002, perinottoetal2004}, depending upon the development of the density enhancements due to the ionization front or the swept-up inner rim.  It is therefore not surprising that there is less difference between the distributions of densities and diameters than there is between the distributions of line widths.  Likewise, since the H$\beta$ luminosity will depend upon how much mass is swept up by the AGB envelope \citep{villaveretal2002}, and that this may vary significantly from one object to the next depending upon location, the H$\beta$ luminosities may be somewhat scattered.  

Our findings confirm those of \citet{heap1993} and \citet{medinaetal2006}, who found correlations between the nebular kinematics and the evolutionary state of the central star.  Our findings, however, are based upon a sample of planetary nebulae that is larger, more homogeneous, and whose analysis is entirely independent of models.  It is perhaps not surprising that our results demonstrate much more clearly the intimate relationship between the central star's evolution and the kinematics of the nebular shell.  

Perhaps, the most relevant studies for comparison are those of \citet{dopitaetal1985, dopitaetal1988} for planetary nebulae in the Magellanic Clouds.  They found that the nebular expansion velocity depends upon the excitation class and H$\beta$ luminosity of the nebula.  Both the excitation class and the H$\beta$ luminosity are functions of the evolutionary stage of both the central star and nebular shell.  In other words, the kinematics of planetary nebulae in the Magellanic Clouds  require a coordination between the evolutionary states of the central star and the nebular shell.  Our results clearly demand the same sort of coordination between the evolution of the central star and the surrounding nebular shell.  

The exact relationship between the central star and the nebular kinematics need not be conserved between the Magellanic Clouds and the Bulge.  Indeed, it is difficult to compare the results directly as many complications may arise.  Ideally, the nebular kinematics should be related to the properties of the central star in both environments (temperature, wind velocity, mass loss rate).  However, for most of the objects in both the Bulge and the Magellanic Clouds, observations of the central star's wind are unavailable.  Excitation classes or line ratios provide indirect temperature diagnostics, but these will usually depend upon the chemical composition of the nebula and are also susceptible to systematic biases if there are differences in the distributions of the central star masses or AGB wind properties.  

Our findings, like those of \citet{gesickietal2006}, do not support the claim that the nebular shells surrounding WR central stars have larger expansion velocities \citep{medinaetal2006}.  We made no special attempt to include planetary nebulae with WR central stars in our sample, so it includes only eight examples (see Table \ref{table_objects}).  Of these, five of the central stars are cool and three are hot enough to produce \ion{He}{2} emission.  With such small numbers, the statistics are clearly not conclusive.  While the cooler WR central stars are found in nebulae with a higher mean velocity than the rest of the planetary nebulae with cool central stars, the hotter WR central stars are found in nebulae with the same line widths as the rest of the sample of hotter central stars.  Perhaps, WR central stars are able to initiate the wind-driven phase of nebular acceleration earlier, but they may not necessarily afford greater acceleration in the long run.  The planetary nebulae with normal central stars in our sample span a larger range in line width than do the expansion velocities observed in nebulae surrounding WR central stars by \citet{medinaetal2006}.  Resolving this issue will require a more careful, dedicated study.  

Therefore, our observations of bright planetary nebulae in the Milky Way bulge confirm the kinematics predicted by theoretical models.  It is notable that observations not only recover the sequence of nebular acceleration that is predicted, but also the coincidence of these stages with particular properties of the central stars.  The details of the models differ considerably in parameters that are difficult to both observe and model, such as the structures within the AGB envelopes or the transition time between the AGB and planetary nebula stages.  

\section{Conclusions}

We have obtained kinematic data for a large sample of planetary nebulae in the Milky Way bulge, selected with the goal of simulating samples of bright extragalactic planetary nebulae in bulge-like environments.  For most of the sample, our criteria also  included a reddening-corrected H$\beta$ flux exceeding $\log F(\mathrm H\beta)>-12.0$\,dex and [\ion{O}{3}]$\lambda 5007/\mathrm H\beta > 6$, in addition to the usual criterion of projected proximity to the Galactic center.  We measure line widths for H$\alpha$ and [\ion{O}{3}]$\lambda 5007$.  Our sample is the largest that has been selected and observed in a homogeneous way.  We have also analyzed it in a completely model-independent manner.  

For half of the sample, the \ion{He}{2}$\lambda 6560$ line is observed in the H$\alpha$ spectra.  Since this line appears only for hotter central stars, it allows us to divide our sample according to the evolutionary stage of the central star, with the more evolved objects found in the subsample containing the hotter central stars.  The kinematics of the two samples are significantly different, with the sample containing the hotter central stars expanding faster.  We also compare the diameters, electron densities (from [\ion{S}{2}]$\lambda\lambda 6716,6731$), and H$\beta$ luminosities for the two subsamples.  In all cases, there are statistically significant differences.  The planetary nebulae hosting the hotter central stars are larger, less dense, and less luminous than their counterparts with cooler central stars.  All of these differences exceed a statistical significance of 99\%.  Also, all of these differences are compatible with the results of hydrodynamical models \citep[e.g.,][]{mellema1994, villaveretal2002, perinottoetal2004}.  

Our primary conclusion is that we have clearly observed the acceleration of the nebular shells in planetary nebulae in the bulge of the Milky Way and that this occurs during the early evolution of their central stars.  Our findings, based upon a large, homogeneous sample, constitute the first unequivocal evidence that the nebular kinematics depend upon the evolutionary state of the central star for planetary nebulae in the Milky Way and are far clearer and more convincing than any previous results.  Hence, there is now very clear evidence that this dependence occurs in at least two environments: the Magellanic Clouds and the bulge of the Milky Way.  Given the differences in the stellar populations in these two environments, it is likely that the dependence of the nebular kinematics on the evolutionary state of central star is more general and that it occurs in all environments.  

\acknowledgments

We thank the technical personnel at the OAN-SPM, and particularly Gabriel Garc\'\i a, Gustavo Melgoza, Salvador Monrroy, and Felipe Montalvo who were the telescope operators during our observing runs.  Their excellent support was a great help in obtaining the data presented here.  We thank S. G\'orny for kindly providing a list of classifications for the central stars of planetary nebulae towards the Galactic bulge.  We acknowledge financial support throughout this project from CONACyT through grants 37214 and 43121 and from UNAM-DGAPA via grants IN112103, IN108406-2, IN108506-2, and IN116908-3.  We thank the anonymous referee for a very constructive report that helped improve this manuscript.

\clearpage

\begin{deluxetable}{lllccccccccll}
\tabletypesize{\scriptsize}
\rotate
\tablecaption{Bulge Planetary Nebula Sample\label{table_objects}} 
\tablewidth{0pt}
\tablehead{
\colhead{Object} & \colhead{PN G} & \colhead{Run} & \colhead{FWHM(H$\alpha$)$^{\mathrm a}$} & \colhead{$\Delta V_{0.5}(\mathrm H\alpha)$} & \colhead{FWHM(5007)$^{\mathrm a}$} & \colhead{$\Delta V_{0.5}(5007)$} & \multicolumn{2}{c}{diameter (in H$\alpha$)$^{\mathrm b}$} & \colhead{$N_e$$^{\mathrm c}$} & \colhead{$\log F(\mathrm H\beta)$} & \colhead{6560?} & \colhead{WR?}$^{\mathrm d}$ \\
\colhead{} & \colhead{} & \colhead{} & \colhead{(\AA)} & \colhead{(km/s)} & \colhead{(\AA)} & \colhead{(km/s)} & \colhead{50\%\,$I_{max}$} & \colhead{10\%\,$I_{max}$} & \colhead{(cm$^{-3}$)} & \colhead{(erg/s)} & \colhead{} & \colhead{}
}
\startdata
Bl 3-13  & 000.9-02.0 & 2006 Jun & $ 0.9427\pm 0.0028 $ & $ 16.05\pm 0.06 $ & $ 0.6063\pm 0.0079 $ & $ 16.93\pm 0.24 $ & 2.4 &  5.5 &  1765 & 34.19 & no  & wels \\
Cn 1-5   & 002.2-09.4 & 2004 Jun & $ 1.0384\pm 0.0026 $ & $ 18.87\pm 0.06 $ & $  1.251\pm  0.041 $ & $ 36.9 \pm 1.2  $ & 4.4 &  8.6 &\ldots & 35.21 & no  & WR   \\
Cn 2-1   & 356.2-04.4 & 2004 Jun & $ 0.9255\pm 0.0039 $ & $ 15.51\pm 0.09 $ & $ 0.6826\pm 0.0074 $ & $ 19.36\pm 0.22 $ & 2.7 &  4.8 &  6967 & 34.97 & no  & no   \\
H 1-1    & 343.4+11.9 & 2004 Jun & $  1.617\pm  0.019 $ & $ 34.04\pm 0.43 $ & $  1.238\pm  0.043 $ & $ 36.5 \pm 1.3  $ & 2.5 &  4.4 &\ldots & 33.88 & yes & no   \\
H 1-11   & 002.6+08.2 & 2006 Jun & $ 0.9219\pm 0.0057 $ & $ 15.40\pm 0.13 $ & $  0.625\pm  0.011 $ & $ 17.53\pm 0.32 $ & 4.4 &  7.3 & 10803 & 34.43 & yes & wels \\
H 1-14   & 001.7+05.7 & 2005 Jul & $  1.803\pm  0.036 $ & $ 38.62\pm 0.81 $ & $  1.313\pm  0.068 $ & $ 38.8 \pm 2.0  $ & 4.3 &  7.4 &   985 & 34.30 & yes & no   \\
H 1-16   & 000.1+04.3 & 2005 May & $  1.026\pm  0.003 $ & $ 18.51\pm 0.07 $ & $  0.743\pm  0.012 $ & $ 21.25\pm 0.36 $ & 2.8 &  4.7 &  7035 & 35.11 & yes & no   \\
H 1-17   & 358.3+03.0 & 2005 Jul & $ 0.9635\pm 0.0035 $ & $ 16.67\pm 0.08 $ & $ 0.6150\pm 0.0072 $ & $ 17.20\pm 0.22 $ & 2.1 &  4.7 & 16190 & 35.29 & no  & no   \\
H 1-18   & 357.6+02.6 & 2004 Jun & $ 0.8105\pm 0.0026 $ & $ 11.67\pm 0.06 $ & $ 0.5062\pm 0.0039 $ & $ 13.66\pm 0.12 $ & 2.6 &  4.8 &  8627 & 34.94 & yes & no   \\
H 1-20   & 358.9+03.2 & 2003 Jun & $ 0.9361\pm 0.0025 $ & $ 15.85\pm 0.06 $ & $ 0.6905\pm 0.0042 $ & $ 19.53\pm 0.13 $ & 2.4 &  4.9 &  3811 & 34.85 & no  & no   \\
H 1-23   & 357.6+01.7 & 2005 May & $ 1.0266\pm 0.0080 $ & $ 18.54\pm 0.18 $ & $  0.754\pm  0.016 $ & $ 21.62\pm 0.49 $ & 3.2 &  5.9 &  3840 & 35.05 & no  & no   \\
H 1-27   & 005.0+04.4 & 2003 Jun & $ 1.0238\pm 0.0029 $ & $ 18.48\pm 0.07 $ & $ 0.7004\pm 0.0075 $ & $ 19.94\pm 0.22 $ & 1.7 &  3.8 & 18385 & 34.83 & no  & no   \\
H 1-30   & 352.0-04.6 & 2006 Jun & $ 0.9616\pm 0.0044 $ & $ 16.62\pm 0.10 $ & $ 0.6000\pm 0.0052 $ & $ 16.73\pm 0.16 $ & 3.2 &  4.4 &  6495 & 34.20 & yes & no   \\
H 1-31   & 355.1-02.9 & 2005 May & $ 0.9932\pm 0.0037 $ & $ 17.56\pm 0.08 $ & $ 0.6646\pm 0.0038 $ & $ 18.79\pm 0.11 $ & 2.8 &  5.1 & 12957 & 34.84 & yes & no   \\
H 1-32   & 355.6-02.7 & 2005 May & $ 0.7685\pm 0.0026 $ & $ 10.08\pm 0.06 $ & $ 0.4385\pm 0.0033 $ & $ 11.37\pm 0.10 $ & 3.1 &  5.2 &  7897 & 34.04 & no  & no   \\
H 1-33   & 355.7-03.0 & 2004 Jun & $ 0.7904\pm 0.0018 $ & $ 10.93\pm 0.04 $ & $ 0.5218\pm 0.0022 $ & $ 14.17\pm 0.07 $ & 3.9 &  6.5 &  3215 & 34.72 & no  & no   \\
H 1-40   & 359.7-02.6 & 2005 May & $  1.014\pm  0.014 $ & $ 18.16\pm 0.32 $ & $  0.643\pm  0.014 $ & $ 18.09\pm 0.42 $ & 3.1 &  7.0 & 11745 & 35.42 & no  & no   \\
H 1-41   & 356.7-04.8 & 2006 Jun & $ 1.1607\pm 0.0091 $ & $ 22.30\pm 0.21 $ & $  0.832\pm  0.014 $ & $ 24.05\pm 0.43 $ & 5.7 &  8.8 &  2525 & 34.35 & yes & wels \\
H 1-42   & 357.2-04.5 & 2006 Jul & $ 0.9351\pm 0.0059 $ & $ 15.81\pm 0.13 $ & $ 0.5362\pm 0.0080 $ & $ 14.65\pm 0.24 $ & 3.0 &  5.2 &  2270 & 34.92 & no  & wels \\
H 1-45   & 002.0-02.0 & 2005 Jul & $  1.590\pm  0.011 $ & $ 33.37\pm 0.26 $ & $  1.016\pm  0.011 $ & $ 29.70\pm 0.34 $ & 2.7 &  4.4 &\ldots & 35.03 & yes & no   \\
H 1-50   & 358.7-05.2 & 2004 Jun & $ 1.0263\pm 0.0018 $ & $ 18.53\pm 0.04 $ & $ 0.7291\pm 0.0038 $ & $ 20.82\pm 0.11 $ & 2.2 &  4.5 &  6920 & 34.95 & yes & no   \\
H 1-54   & 002.1-04.2 & 2007 Aug & $ 0.9029\pm 0.0045 $ & $ 14.81\pm 0.10 $ & $ 0.5304\pm 0.0028 $ & $ 14.50\pm 0.08 $ & 2.8 &  5.2 & 11594 & 34.78 & no  & no   \\
H 1-56   & 001.7-04.6 & 2007 Aug & $ 0.8261\pm 0.0013 $ & $ 12.25\pm 0.03 $ & $ 0.5650\pm 0.0048 $ & $ 15.67\pm 0.14 $ & 3.2 &  6.1 &  1164 & 34.52 & no  & wels \\
H 1-59   & 003.8-04.3 & 2005 May & $  1.237\pm  0.015 $ & $ 24.33\pm 0.34 $ & $  1.115\pm  0.033 $ & $ 32.76\pm 0.99 $ & 3.6 &  6.8 &  1100 & 34.26 & yes & no   \\
H 1-60   & 004.2-04.3 & 2005 May & $ 0.9560\pm 0.0076 $ & $ 16.45\pm 0.17 $ & $  0.672\pm  0.013 $ & $ 19.04\pm 0.38 $ & 3.2 &  7.0 &\ldots & 34.26 & no  & wels \\
H 1-67   & 009.8-04.6 & 2005 Jul & $  1.474\pm  0.030 $ & $ 30.46\pm 0.68 $ & $      \ldots      $ & $     \ldots    $ & 5.2 &  9.5 &  1455 & 34.36 & yes & WR   \\
H 2-10   & 358.2+03.5 & 2004 Jun & $ 1.1028\pm 0.0093 $ & $ 20.70\pm 0.21 $ & $ 0.7907\pm 0.0038 $ & $ 22.75\pm 0.11 $ & 1.9 &  4.1 &  8140 & 34.61 & no  & no   \\
H 2-11   & 000.7+04.7 & 2005 Jul & $ 0.7697\pm 0.0058 $ & $ 10.13\pm 0.13 $ & $ 0.4684\pm 0.0055 $ & $ 12.39\pm 0.16 $ & 2.6 &  5.2 & 15550 & 35.03 & no  & wels \\
H 2-18   & 006.3+04.4 & 2004 Jun & $  1.473\pm  0.020 $ & $ 30.44\pm 0.47 $ & $  1.118\pm  0.028 $ & $ 32.84\pm 0.85 $ & 2.9 &  6.6 &\ldots & 34.14 & yes & no   \\
Hb 8     & 003.8-17.1 & 2004 Jun & $ 0.9662\pm 0.0022 $ & $ 16.75\pm 0.05 $ & $ 0.6465\pm 0.0037 $ & $ 18.21\pm 0.11 $ & 1.8 &  4.1 & 14420 & 34.05 & no  & no   \\
He 2-250 & 000.7+03.2 & 2003 Jun & $  1.185\pm  0.021 $ & $ 22.96\pm 0.48 $ & $  0.797\pm  0.013 $ & $ 22.98\pm 0.40 $ & 3.6 &  7.9 &  2430 & 34.62 & yes & no   \\
Hf 2-1   & 355.4-04.0 & 2005 May & $  1.774\pm  0.059 $ & $ 37.9 \pm 1.3  $ & $  1.285\pm  0.069 $ & $ 37.9 \pm 2.1  $ & 9.1 & 19.7 &\ldots & 34.17 & yes & no   \\
K 5-1    & 000.4+04.4 & 2006 Jul & $ 0.8709\pm 0.0027 $ & $ 13.76\pm 0.06 $ & $ 0.6191\pm 0.0078 $ & $ 17.34\pm 0.23 $ & 3.6 &  5.7 &  1400 & 34.61 & no  & wels \\
K 5-3    & 002.6+05.5 & 2006 Jul & $  1.368\pm  0.041 $ & $ 27.77\pm 0.94 $ & $  0.538\pm  0.020 $ & $ 14.70\pm 0.59 $ & 2.6 &  9.5 &  1300 & 34.01 & yes & WR   \\
K 5-4    & 351.9-01.9 & 2006 Jul & $ 0.8187\pm 0.0013 $ & $ 11.97\pm 0.03 $ & $ 0.4946\pm 0.0016 $ & $ 13.28\pm 0.05 $ & 2.2 &  3.4 &  7600 & 34.99 & no  & no   \\
K 5-5    & 001.5+03.6 & 2006 Jul & $ 0.9265\pm 0.0047 $ & $ 15.54\pm 0.11 $ & $  0.612\pm  0.011 $ & $ 17.12\pm 0.31 $ & 2.5 &  4.1 & 10900 & 34.85 & no  & no   \\
K 5-6    & 003.6+04.9 & 2006 Jul & $  1.302\pm  0.018 $ & $ 26.04\pm 0.42 $ & $  0.866\pm  0.027 $ & $ 25.09\pm 0.82 $ & 3.6 &  8.6 &  1000 & 33.85 & yes & no   \\
K 5-7    & 003.1+04.1 & 2006 Jul & $  1.619\pm  0.023 $ & $ 32.49\pm 0.52 $ & $  1.323\pm  0.035 $ & $ 39.1 \pm 1.0  $ & 8.9 & 13.2 &   300 & 33.62 & yes & no   \\
K 5-9    & 355.54-1.4 & 2006 Jul & $  1.289\pm  0.035 $ & $ 25.72\pm 0.81 $ & $  0.964\pm  0.043 $ & $ 28.1 \pm 1.3  $ & 4.0 &  7.7 &  3500 & 34.60 & yes & no   \\
K 5-11   & 002.3+02.2 & 2006 Jul & $ 0.9577\pm 0.0054 $ & $ 16.51\pm 0.12 $ & $ 0.6465\pm 0.0087 $ & $ 18.22\pm 0.26 $ & 5.6 & 10.8 &   604 & 34.31 & no  & no   \\
K 5-12   & 353.5-03.3 & 2006 Jul & $  1.548\pm  0.017 $ & $ 32.33\pm 0.38 $ & $  1.208\pm  0.031 $ & $ 35.59\pm 0.93 $ & 4.1 &  7.3 &  4480 & 34.07 & yes & no   \\
K 5-14   & 003.9+02.6 & 2007 Aug & $ 1.0314\pm 0.0087 $ & $ 18.69\pm 0.20 $ & $ 0.5986\pm 0.0052 $ & $ 16.50\pm 0.16 $ & 2.0 &  3.6 &  8000 & 34.16 & yes & no   \\
K 5-17   & 004.3+02.1 & 2007 Aug & $  1.435\pm  0.027 $ & $ 29.48\pm 0.62 $ & $  1.043\pm  0.037 $ & $ 30.5 \pm 1.1  $ & 2.4 &  5.1 &  9960 & 34.41 & yes & no   \\
K 5-19   & 005.1+02.0 & 2007 Aug & $  1.409\pm  0.016 $ & $ 28.81\pm 0.36 $ & $  1.191\pm  0.030 $ & $ 35.08\pm 0.91 $ & 3.0 &  5.9 &  1170 & 34.15 & yes & no   \\
K 5-20   & 356.8-03.0 & 2007 Aug & $ 0.9118\pm 0.0054 $ & $ 15.09\pm 0.12 $ & $  0.639\pm  0.010 $ & $ 18.02\pm 0.30 $ & 3.2 &  6.1 &   498 & 34.00 & no  & no   \\
M 1-19   & 351.1+04.8 & 2005 May & $ 0.8781\pm 0.0022 $ & $ 14.00\pm 0.05 $ & $ 0.5640\pm 0.0056 $ & $ 15.56\pm 0.17 $ & 2.6 &  5.4 &  6370 & 34.99 & no  & wels \\
M 1-20   & 006.1+08.3 & 2004 Jun & $ 0.7527\pm 0.0037 $ & $  9.43\pm 0.08 $ & $ 0.2969\pm 0.0031 $ & $  5.97\pm 0.09 $ & 3.0 &  4.5 &  7500 & 35.07 & no  & wels \\
M 1-29   & 359.1-01.7 & 2004 Jun & $ 1.0319\pm 0.0033 $ & $ 18.69\pm 0.08 $ & $ 0.7693\pm 0.0060 $ & $ 22.08\pm 0.18 $ & 5.3 &  8.8 &  3674 & 35.34 & yes & wels \\
M 1-31   & 006.4+02.0 & 2005 Jul & $ 0.7954\pm 0.0044 $ & $ 11.11\pm 0.10 $ & $ 0.4899\pm 0.0025 $ & $ 13.11\pm 0.07 $ & 2.2 &  4.3 &  7830 & 35.32 & no  & wels \\
M 1-35   & 003.9-02.3 & 2007 Aug & $ 1.0215\pm 0.0063 $ & $ 18.40\pm 0.14 $ & $ 0.7336\pm 0.0073 $ & $ 21.03\pm 0.22 $ & 4.0 &  6.7 &  6040 & 35.10 & no  & no   \\
M 1-42   & 002.7-04.8 & 2003 Jun & $ 1.0485\pm 0.0038 $ & $ 19.18\pm 0.09 $ & $ 0.6772\pm 0.0055 $ & $ 19.21\pm 0.16 $ & 7.9 & 12.4 &  1287 & 34.95 & yes & no   \\
M 1-48   & 013.4-03.9 & 2005 Jul & $ 0.8262\pm 0.0024 $ & $ 12.23\pm 0.05 $ & $ 0.5301\pm 0.0046 $ & $ 14.45\pm 0.14 $ & 3.8 &  6.9 &  1525 & 34.06 & yes & no   \\
M 2-13   & 011.1+11.5 & 2006 Jun & $ 0.7047\pm 0.0025 $ & $  7.25\pm 0.06 $ & $ 0.4008\pm 0.0020 $ & $ 10.04\pm 0.06 $ & 2.8 &  5.6 &  4185 & 34.26 & no  & no   \\
M 2-15   & 011.0+06.2 & 2006 Jun & $  1.271\pm  0.028 $ & $ 25.23\pm 0.64 $ & $  0.937\pm  0.035 $ & $ 27.3 \pm 1.0  $ & 6.6 &  9.1 &  2130 & 34.57 & yes & no   \\
M 2-16   & 357.4-03.2 & 2004 Jun & $ 1.1090\pm 0.0057 $ & $ 20.87\pm 0.13 $ & $ 0.7905\pm 0.0077 $ & $ 22.74\pm 0.23 $ & 2.5 &  4.9 &  4625 & 35.07 & yes & no   \\
M 2-20   & 000.4-01.9 & 2006 Jul & $ 1.2428\pm 0.0041 $ & $ 24.50\pm 0.09 $ & $ 0.8262\pm 0.0027 $ & $ 23.86\pm 0.08 $ & 2.1 &  4.3 &  2023 & 35.02 & no  & WR   \\
M 2-21   & 000.7-02.7 & 2005 May & $  1.280\pm  0.016 $ & $ 25.49\pm 0.37 $ & $  0.945\pm  0.026 $ & $ 27.52\pm 0.78 $ & 2.8 &  5.3 &    10 & 34.65 & yes & wels \\
M 2-22   & 357.4-04.6 & 2007 Aug & $  1.234\pm  0.030 $ & $ 24.27\pm 0.67 $ & $  0.903\pm  0.039 $ & $ 26.3 \pm 1.1  $ & 3.1 &  6.9 &  2300 & 34.18 & yes & no   \\
M 2-23   & 002.2-02.7 & 2004 Jun & $  0.855\pm 0.0021 $ & $ 13.22\pm 0.05 $ & $ 0.5258\pm 0.0030 $ & $ 14.30\pm 0.09 $ & 2.5 &  4.3 & 10600 & 35.00 & no  & no   \\
M 2-26   & 003.6-02.3 & 2006 Jul & $  1.319\pm  0.024 $ & $ 26.50\pm 0.56 $ & $  1.011\pm  0.033 $ & $ 29.55\pm 0.99 $ & 8.2 & 11.2 &   452 & 34.14 & yes & no   \\
M 2-27   & 359.9-04.5 & 2004 Jun & $ 0.9625\pm 0.0034 $ & $ 16.64\pm 0.08 $ & $ 0.6230\pm 0.0044 $ & $ 17.46\pm 0.13 $ & 1.9 &  6.4 &  7360 & 35.11 & no  & wels \\
M 2-29   & 004.0-03.0 & 2004 Jun & $ 0.7864\pm 0.0026 $ & $ 10.77\pm 0.06 $ & $ 0.4771\pm 0.0036 $ & $ 12.68\pm 0.11 $ & 3.6 &  7.1 &  2670 & 34.53 & no  & no   \\
M 2-30   & 003.7-04.6 & 2004 Jun & $  1.184\pm  0.011 $ & $ 22.91\pm 0.25 $ & $  0.864\pm  0.016 $ & $ 25.02\pm 0.47 $ & 3.2 &  5.6 &  2268 & 34.83 & yes & wels \\
M 2-31   & 006.0-03.6 & 2004 Jun & $ 1.1907\pm 0.0068 $ & $ 23.10\pm 0.16 $ & $ 0.8266\pm 0.0060 $ & $ 23.87\pm 0.18 $ & 2.3 &  5.6 &  4710 & 35.19 & no  & WR   \\
M 2-33   & 002.0-06.2 & 2007 Aug & $ 0.6988\pm 0.0031 $ & $  6.98\pm 0.07 $ & $ 0.3441\pm 0.0046 $ & $  7.98\pm 0.14 $ & 4.1 &  6.1 &  1259 & 34.94 & no  & wels \\
M 2-39   & 008.1-04.7 & 2006 Jul & $ 1.0341\pm 0.0053 $ & $ 18.76\pm 0.12 $ & $ 0.5258\pm 0.0063 $ & $ 14.31\pm 0.19 $ & 2.4 &  3.6 &  3165 & 34.62 & no  & wels \\
M 2-4    & 349.8+04.4 & 2007 Aug & $ 0.8088\pm 0.0041 $ & $ 11.63\pm 0.09 $ & $ 0.5025\pm 0.0035 $ & $ 13.58\pm 0.10 $ & 1.7 &  4.4 &  6970 & 35.39 & no  & no   \\
M 2-8    & 352.1+05.1 & 2006 Jun & $ 0.9742\pm 0.0019 $ & $ 17.00\pm 0.04 $ & $ 0.6275\pm 0.0021 $ & $ 17.61\pm 0.06 $ & 2.2 &  5.3 &  1670 & 34.44 & yes & WR   \\
M 3-10   & 358.2+03.6 & 2004 Jun & $ 1.0963\pm 0.0085 $ & $ 20.52\pm 0.19 $ & $ 0.7351\pm 0.0069 $ & $ 21.01\pm 0.21 $ & 2.5 &  4.7 & 11497 & 34.99 & yes & no   \\
M 3-14   & 355.4-02.4 & 2004 Jun & $ 1.1585\pm 0.0075 $ & $ 22.23\pm 0.17 $ & $ 0.8005\pm 0.0087 $ & $ 23.06\pm 0.26 $ & 3.5 &  7.5 &  3180 & 35.28 & yes & no   \\
M 3-15   & 006.8+04.1 & 2004 Jun & $ 0.9783\pm 0.0033 $ & $ 17.12\pm 0.08 $ & $ 0.6470\pm 0.0062 $ & $ 18.23\pm 0.19 $ & 4.0 &  6.1 &  9450 & 35.67 & no  & WR   \\
M 3-16   & 359.1-02.3 & 2005 May & $ 1.0279\pm 0.0074 $ & $ 18.58\pm 0.17 $ & $ 0.6229\pm 0.0085 $ & $ 17.46\pm 0.25 $ & 6.0 & 11.2 &  1155 & 35.06 & no  & no   \\
M 3-20   & 002.1-02.2 & 2007 Aug & $ 1.1206\pm 0.0093 $ & $ 21.20\pm 0.21 $ & $ 0.7629\pm 0.0046 $ & $ 21.89\pm 0.14 $ & 3.5 &  4.8 &    10 & 34.64 & no  & wels \\
M 3-21   & 355.1-06.9 & 2004 Jun & $ 0.8936\pm 0.0062 $ & $ 14.49\pm 0.14 $ & $ 0.5682\pm 0.0056 $ & $ 15.69\pm 0.17 $ & 1.8 &  4.4 &  3172 & 35.06 & no  & no   \\
M 3-26$^{\mathrm e}$ & 004.8-05.0 & 2005 Sep & $  1.462\pm  0.052 $ & $ 30.2 \pm 1.2  $ & $  1.141           $ & $ 33.5          $ & 7.0 & 10.7 &  1415 & 34.13 & yes & no   \\
M 3-32   & 009.4-09.8 & 2005 Jul & $  1.176\pm  0.012 $ & $ 22.70\pm 0.27 $ & $  0.811\pm  0.013 $ & $ 23.37\pm 0.38 $ & 4.0 &  7.4 &  1850 & 34.38 & yes & no   \\
M 3-33   & 009.6-10.6 & 2004 Jun & $  1.098\pm  0.012 $ & $ 20.55\pm 0.28 $ & $  0.806\pm  0.020 $ & $ 23.24\pm 0.61 $ & 5.3 &  8.8 &  1140 & 34.05 & yes & wels \\
M 3-38   & 356.9+04.4 & 2004 Jun & $ 0.9558\pm 0.0086 $ & $ 16.44\pm 0.20 $ & $ 0.5289\pm 0.0077 $ & $ 14.41\pm 0.23 $ & 3.2 &  4.4 &  9550 & 34.31 & yes & no   \\
M 3-42   & 357.5+03.2 & 2003 Jun & $  1.710\pm  0.096 $ & $ 36.4 \pm 2.2  $ & $  0.602\pm  0.046 $ & $ 16.7 \pm 1.4  $ & 4.1 &  9.8 &  1000 & 33.72 & yes & no   \\
M 3-45   & 359.7-01.8 & 2005 Jul & $ 1.1658\pm 0.0068 $ & $ 22.43\pm 0.16 $ & $ 0.8117\pm 0.0095 $ & $ 23.41\pm 0.28 $ & 5.1 &  7.4 &  7897 & 34.86 & yes & no   \\
M 3-54   & 018.6-02.2 & 2006 Jul & $  1.340\pm  0.020 $ & $ 27.03\pm 0.46 $ & $  0.957\pm  0.026 $ & $ 27.89\pm 0.79 $ & 4.4 &  7.4 &  1385 & 33.97 & yes & no   \\
M 4-3    & 357.2+07.4 & 2005 May & $ 0.8502\pm 0.0022 $ & $ 13.07\pm 0.05 $ & $ 0.5271\pm 0.0025 $ & $ 14.35\pm 0.07 $ & 3.6 &  6.0 &\ldots & 35.27 & no  & no   \\
M 4-6    & 358.6+01.8 & 2004 Jun & $ 1.1311\pm 0.0067 $ & $ 21.48\pm 0.15 $ & $  0.772\pm  0.014 $ & $ 22.15\pm 0.40 $ & 3.7 &  6.6 & 12881 & 34.26 & no  & no   \\
M 4-7    & 358.5-02.5 & 2006 Jun & $ 1.0558\pm 0.0060 $ & $ 19.38\pm 0.14 $ & $  0.730\pm  0.012 $ & $ 20.86\pm 0.36 $ & 4.8 &  8.0 &  1580 & 34.15 & no  & no   \\
PC 12    & 000.1+17.2 & 2005 May & $ 0.8346\pm 0.0019 $ & $ 12.53\pm 0.04 $ & $ 0.4749\pm 0.0019 $ & $ 12.61\pm 0.06 $ & 3.1 &  6.3 &  5110 & 34.27 & no  & no   \\
Te 1580  & 002.6+02.1 & 2007 Aug & $  1.883\pm  0.065 $ & $ 40.6 \pm 1.5  $ & $  1.367\pm  0.075 $ & $ 40.42\pm 2.23 $ & 7.4 &  9.9 &  1200 & 34.48 & yes & no   \\
\enddata
\tablenotetext{a}{This is the observed line width, uncorrected for instrumental, thermal, or fine structure broadening.}
\tablenotetext{b}{These diameters were measured by collapsing the H$\alpha$ spectra into one-dimensional spatial profiles, and measuring the diameters at 50\% and 10\% of peak intensity.}
\tablenotetext{c}{The electron densities are measured from the [\ion{S}{2}]$\lambda\lambda 6716,6731$ lines.  When no density is given, this information is unavailable.}
\tablenotetext{d}{Classifications of \lq\lq WR" and \lq\lq wels" are taken from an extended version of Table 4 from \citet{gornyetal2004} kindly provided by S. G\'orny.  A classification of \lq\lq no" indicates that no information is available or that the central star is classified as non-WR.}
\tablenotetext{e}{The [\ion{O}{3}]$\lambda$5007 line width was measured using IRAF's splot.}
\end{deluxetable}

\begin{deluxetable}{lll}
\tabletypesize{\scriptsize}
\tablecaption{Statistical Tests (U-test)\label{table_statistics}} 
\tablewidth{0pt}
\tablehead{
\colhead{Hypothesis} & 
\colhead{Prob(False)} & \colhead{Comment}
}
\startdata
$\Delta V_{0.5, 5007}(\mathrm{hotter\ CS})>\Delta V_{0.5, 5007}(\mathrm{cooler\ CS}$) & $1.3\times 10^{-7}$ & H 1-67 excluded \\
$\Delta V_{0.5, \mathrm H\alpha}(\mathrm{hotter\ CS})>\Delta V_{0.5, \mathrm H\alpha}(\mathrm{cooler\ CS})$ & $7.4\times 10^{-10}$ & \\
diameter(hotter CS) $>$ diameter(cooler CS) & $7.0\times 10^{-4}$ & measured at 10\% intensity \\
density(cooler CS) $>$ density(hotter CS) & $8.0\times 10^{-3}$ & based upon[S~\sc{ii}]$\lambda\lambda 6716,6731$ \\
$L(\mathrm H\beta, \mathrm{cooler\ CS})>L(\mathrm H\beta, \mathrm{hotter\ CS})$ & $2.2\times 10^{-4}$ \\
\enddata
\end{deluxetable}

\clearpage

\begin{thebibliography}{}

\bibitem[Aller \& Keyes (1987)]{allerkeyes1987} Aller, L. H., \& Keyes, C. D. 1987, \apjs, 65, 405

\bibitem[Breitschwerdt \& Kahn (1990)]{breitschwerdtkahn1990} Breitschwerdt, D., \& Kahn, F. D. 1990, \mnras, 244, 521

\bibitem[Chevalier (1997)]{chevalier1997} Chevalier, R. A. 1997, \apj, 488, 263

\bibitem[Cuisinier et al. (1996)]{cuisinieretal1996} Cuisinier, F., Acker, A., \& K\"oppen, J. 1996, \aap, 307, 215

\bibitem[Cuisinier et al. (2000)]{cuisinieretal2000} Cuisinier, F., Maciel, W. J., K\"oppen, J., Acker, A., \& Stenholm, B. 2000, A\&A, 353, 543

\bibitem[Dopita et al. (1985)]{dopitaetal1985} Dopita, M. A., Ford, H. C., Lawrence, C. J., \& Webster, B. L. 1985, \apj, 296, 390

\bibitem[Dopita et al. (1988)]{dopitaetal1988} Dopita, M. A., Meatheringham, S. J., Webster, B. L., \& Ford, H. C. 1988, ApJ, 327, 639

\bibitem[Dudziak et al. (2000)]{dudziaketal2000} Dudziak, G., P\'equignot, D., Zijlstraa, A. A., \& Walsh, J. R. 2000, A\&A, 363, 717

\bibitem[Escudero \& Costa (2001)]{escuderocosta2001} Escudero, A. V., \& Costa, R. D. D. 2001, A\&A, 380, 300

\bibitem[Escudero et al. (2004)]{escuderoetal2004} Escudero, A. V., Costa, R. D. D., \& Maciel, W. J. 2004, A\&A, 414, 211

\bibitem[Exter et al. (2004)]{exteretal2004} Exter, K. M., Barlow, M. J., \& Walton, N. A. 2004, MNRAS, 349, 1291

\bibitem[Garc\'\i a-Diaz et al. (2008)]{garciadiazetal2008} Garc\'\i a-D\'\i az, Ma. T., Henney, W. J., L\'opez, J. A., \& Doi, T. 2008, \rmxaa, 44, 181

\bibitem[Garc\'\i a-Segura et al. (2006)]{garciaseguraetal2006} Garc\'\i a-Segura, G., L\'opez, J. A., Steffen, W., Meaburn, J., \& Manchado, A. 2006, \apjl, 646, 61

\bibitem[Gesicki \& Zijlstra (2000)]{gesickizijlstra2000} Gesicki, K., \& Zijlstra, A. A. 2000, \aap, 358, 1058

\bibitem[Gesicki et al. (2006)]{gesickietal2006} Gesicki, K., Zijlstra, A. A., Acker, A., G\'orny, S. K., Godziewski, K., \& Walsh, J. R. 2006, \aap, 451, 925

\bibitem[Gon\c calves et al. (2007)]{goncalvesetal2007} Gon\c calves, D. R., Magrini, L., Leisy, P., \& Corradi, R. L. M. 2007, MNRAS, 375, 715

\bibitem[G\'orny et al. (2004)]{gornyetal2004} G\'orny, S. K., Stasi\'nska, G., Escudero, A. V., \& Costa, R. D. D. 2004, A\&A, 427, 231

\bibitem[Heap (1993)]{heap1993} Heap, S. R., in IAU Symp. 155: Planetary Nebulae, eds. R. Weinberger \& A. Acker (Reidel Publishing: Dordrecht: the Netherlands), 23

\bibitem[Jacoby \& Ciardullo (1999)]{jacobyciardullo1999} Jacoby, G. H., \& Ciardullo, R. 1999, ApJ, 515, 169

\bibitem[Kahn \& West (1985)]{kahnwest1985} Kahn, F. D., \& West, K. A. 1985, \mnras, 212, 837

\bibitem[Kahn \& Breitschwerdt (1990)]{kahnbreitschwerdt1990} Kahn, F. D., \& Breitschwerdt, D. 1990, \mnras, 242, 505

\bibitem[Kwok et al. (1978)]{kwoketal1978} Kwok, S., Purton, C. R., \& Fitzgerald, P. M. 1978, \apjl, 219, 125

\bibitem[Lang (1980)]{lang1980} Lang, K. R. 1980, Astrophysical Formulae (Springer-Verlag: Berlin, Heidelberg)

\bibitem[Lewis (1991)]{lewis1991} Lewis, B. M. 1991, \aj, 101, 254

\bibitem [Massey et al. (1992)]{masseyetal1992} Massey, P., Valdes, F., \& Barnes, J. 1992, A User's Guide to Reducing Slit Spectra with IRAF, IRAF User Guide, Vol. 2B (Tucson: National Optical Astronomy Observatory)

\bibitem [McCall et al (1985)]{mccalletal1985} McCall, M. L., Rybski, P. M., \& Shields, G. A. 1985, ApJS, 57, 1

\bibitem[Meaburn et al. (1984)]{meaburnetal1984} Meaburn, J., Blundell, B.,
Carling, R., Gregory, D. F., Keir, D., et al. 1984, \mnras, 210, 463

\bibitem[Meaburn et al. (2003)]{meaburnetal2003} Meaburn, J., L\'opez, J. A., Guti\'errez, L., Quiroz, F., Murillo, J. M., et al. 2003, \rmxaa, 39, 185

\bibitem[Medina et al. (2006)]{medinaetal2006} Medina, S., Pe\~na, M., Morisset, C., \& Stasi\'nska, G. 2006, \rmxaa, 42, 53

\bibitem[Mellema (1994)]{mellema1994} Mellema, G. 1994, \aap, 290, 915

\bibitem[M\'endez et al. (2005)]{mendezetal2005} M\'endez, R. H., Thomas, D., Saglia, R. P., Maraston, C., Kudritski, R. P., \& Bender, R. 2005, ApJ, 627, 767

\bibitem[Osterbrock (1989)]{osterbrock1989} Osterbrock, D. E. 1989, Astrophysics of Gaseous Nebulae and Active Galactic Nuclei (Mill Valley, USA: University Science Books)

\bibitem[Perinotto et al. (2004)]{perinottoetal2004} Perinotto, M., Sch\"onberner, D., Steffen, M., \& Calonaci, C. 2004, \aap, 414, 993

\bibitem[Ramstedt et al. (2006)]{ramstedtetal2006} Ramstedt, S., Sch\"oier, F. L., Olofsson, H., \& Lundgren, A. A. 2006, \aap, 454, L103

\bibitem[Ratag et al. (1997)]{ratagetal1997} Ratag, M. A., Pottasch, S. R., Dennefeld, M., \& Menzies, J. 1997, A\&AS, 126, 297

\bibitem[Richer et al. (2009)]{richeretal2009} Richer, M. G., B\'aez, S.-H., L\'opez, J. A., \& Riesgo, H., in preparation

\bibitem [Richer \& McCall (1995)]{richermccall1995} Richer, M. G., \& McCall, M. L. 1995, ApJ, 445, 642

\bibitem[Richer \& McCall (2008)]{richermccall2008} Richer, M. G., \& McCall, M. L. 2008, \apj, accepted; also astro-ph/0805.3669

\bibitem[Richer et al. (1999)]{richeretal1999} Richer, M. G., Stasi\'nska, G., \& McCall, M. L. 1999, A\&AS, 135, 203

\bibitem[Roth et al. (2004)]{rothetal2004} Roth, M. M., Becker, T., Kelz, A., \& Schmoll, J. 2004, ApJ, 603, 531

\bibitem[Rozas et al. (2007)]{rozasetal2007} Rozas, M., Richer, M. G., Steffen, W., Garc\'\i a-Segura, G., \& L\'opez, J. A. 2007, \aap, 467, 603

\bibitem[Sch\"onberner et al. (2007)]{schonberneretal2007} Sch\"onberner, D., Jacob, R., Steffen, M., \& Sandin, C. 2007, \aap, 473, 467

\bibitem[Villaver et al. (2002)]{villaveretal2002} Villaver, E., Manchado, A., \& Garc\'\i a-Segura, G. 2002, \apj, 581, 1204

\bibitem[Wall \& Jenkins (2003)]{walljenkins2003} Wall, J. V., \& Jenkins, C. R. 2003, Practical Statistics for Astronomers (Cambridge University Press: Cambridge, U.K.)

\bibitem[Walsh et al. (1999)]{walshetal1999} Walsh, J. R., Walton, N. A., Jacoby, G. H., \& Peletier, R. F. 1999, A\&A, 346, 753

\bibitem[Webster (1988)]{webster1988} Webster, B. L. 1988, \mnras, 230, 377

\bibitem[Zijlstraa et al. (2006)]{zijlstraaetal2006} Zijlstraa, A. A., Gesicki, K., Walsh, J. R., P\'equignot, D., van Hoof, P. A. M., \& Minniti, D. 2006, \mnras, 369, 875

\end{thebibliography}
\end{document}